\documentclass[floatfix,superscriptaddress,showpacs,amssymb,10pt,aps,prd,reprint,longbibliography]{revtex4-1}

\usepackage{graphicx,epsfig,amssymb} 
\usepackage{amsmath,amsfonts,times}
\usepackage{bm} 

\usepackage[caption=false]{subfig}
\usepackage[usenames]{color}     
\usepackage{natbib}
\usepackage{soul}

\usepackage[linktocpage,colorlinks]{hyperref}

\definecolor{coolblack}{rgb}{0.0, 0.18, 0.39}
\definecolor{darkred}{rgb}{0.5,0,0}
\definecolor{darkgreen}{rgb}{0,0.5,0}
\definecolor{darkblue}{rgb}{0,0,0.5}
\definecolor{lapislazuli}{rgb}{0.15, 0.38, 0.61}
\definecolor{venetianred}{rgb}{0.78, 0.03, 0.08}
\definecolor{bleudefrance}{rgb}{0.19, 0.55, 0.91}
\definecolor{dogwoodrose}{rgb}{0.84, 0.09, 0.41}
\hypersetup{colorlinks=true, citecolor=darkgreen, linkcolor=darkblue, 
	urlcolor = blue}

\newcommand{\dd}{\mathrm{d}}

\begin{document}
	\author{Guanming Li}
	\affiliation{School of Physics and Technology, Wuhan University, Wuhan, 430072, China}
	\author{Zhongze Li}
	\affiliation{School of Physics and Technology, Wuhan University, Wuhan, 430072, China}
	
	\author{Junji Jia}
	\email[Corresponding author:~]{junjijia@whu.edu.cn}
	\affiliation{Department of Astronomy $\&$ MOE Key Laboratory of Artificial Micro- and Nano-structures, School of Physics and Technology, Wuhan University, Wuhan, 430072, China}
	\title{Light deflection in general static and spherically symmetric spacetime with a homogeneous plasma}

	\begin{abstract}
		We developed in this work a perturbative technique to compute the deflection angle of light rays in general static and spherically symmetric spacetimes with a homogeneous non-magnetic plasma. The deflection angle is expressed as a power series of $M/b$ with $M$ and $b$ being the spacetime mass and the impact parameter of the ray. The series coefficients are polynomials of the asymptotic expansion coefficients of the metric functions and the reciprocal of the asymptotic refractive index. When the plasma is dilute, the deflection angle can also be expressed as a dual series of $M/b$ and the frequency ratio between the electron plasma frequency and the asymptotic photon frequency. The series result reveals that for general SSS spacetime, the plasma at the leading order always enhances the deflection angle and therefore increases the apparent angles of the gravitational lensing images. These series results of the deflection angle are then shown to have excellent agreement with those obtained using numerical integration. The general formula of the deflection angle is then applied to the Reissner-Nordstr\"{o}m, charged Horndeski and charged Galileon spacetimes. The effect of the plasma and characteristic parameter of the spacetimes on the deflection angles in these spacetimes was briefly discussed. In the appendix, the deflection angles in previously attempted spacetimes are re-computed and compared with the literature. 
	\end{abstract}
	
	
	\maketitle

\section{Introduction}

The deflection of photons' trajectories in curved spacetime was one of the most famous effects of General Relativity (GR). It not only helped the establishment of GR as the more accurate description of gravity, but also has evolved into gravitational lensing (GL), a powerful tool in astronomy. GL has been used to  find exoplanets \cite{Accioly:2002ck,Accioly:2004bm}, to determine the mass profile of the lens including the dark matter distribution \cite{Dyson:1920cwa,Bhadra:2006fv,Tsupko:2014wza}, and to test gravitational theories beyond GR \cite{He:2016vxc,He:2017fov}. 

The deflection of the light rays is affected by not only the spacetime structure through the gravitational channel, but also its interaction with the contents within the spacetime through the electromagnetic channel. One typical example of such content is the widely existing astronomical plasma. Such plasma exists within galaxies \cite{Mathis:2000qc}, around galaxy clusters \cite{Markevitch:2007dn,Bambi:2024fdo} and even in the observable universe \cite{McQuinn:2015icp}.  The interaction of the photon with the charged medium in the plasma effectively changes the group velocity of the photon and the local refractive index it experiences. These effectively influence the bending of light rays passing through a gravitational lens. 

Many works have been devoted to the study of the deflection of light rays in spacetimes with plasma distributions in recent years. Various methods, including Gauss-Bonnet theorem based methods \cite{Javed:2019ynm,Javed:2021ymu,Ali:2025ymm} and ray tracing \cite{Rogers:2015dla}, have been used for this purpose. Deflection angles in Schwarzschild \cite{perlick2000book,Bisnovatyi-Kogan:2008qbk,Bisnovatyi-Kogan:2010flt}, Reissner-Nordstr\"{o}m (RN) black hole (BH) \cite{Ali:2025ymm} and many other spacetimes \cite{Morozova:2013uyv,Javed:2019ynm, Javed:2021ymu,atamurotovWeakGravitationalLensing2021,Ali:2023eqj,Alimova:2024bjd}, as well as in the strong deflection limit \cite{Tsupko:2013cqa,Atamurotov:2015nra,Perlick:2015vta,Atamurotov:2022iwj}, and with different kinds of plasma distribution \cite{Bisnovatyi-Kogan:2008qbk,Atamurotov:2022iwj} have been studied. On the observational side, the effect of the plasma on the time delay in the strong lens system has been investigated in Ref. \cite{Er:2013efa,Er:2022lad}.

Previously, we have developed a perturbative technique to compute the deflection angle of both null and massive rays \cite{Jia:2020qzt,Huang:2020trl,Jia:2020xbc,Duan:2020tsq,Liu:2025brb}. This method works for both neutral signals and charged particles experiencing both gravitational and electromagnetic interaction \cite{Xu:2021rld,Zhou:2022dze}, and can take the finite distance effect into account naturally \cite{Huang:2020trl,Duan:2023gvm}. It is also applicable to the computation of the time delay in GL \cite{Liu:2020mkf,Liu:2020wcu,Liu:2021ckg}, and works in the strong deflection limit as well \cite{Zhou:2022dze,Duan:2023gvm}. In this work, we generalize this method to the computation of the weak deflection angle of light rays in arbitrary static and spherically symmetric (SSS) spacetimes with a non-magnetized and pressureless plasma. We will show that deflection takes a series form of $M/b$ where $M$ and $b$ are respectively the mass of the spacetime and impact parameter of the trajectory. The coefficients of the series are expressed as a polynomial of the asymptotic expansion coefficients of the metric function and $1/n_\infty$, where $n_\infty$ is the asymptotic refractive index of the photon. When the plasma is dilute, the deflection can be written as a dual series of $M/b$ and the frequency ratio $\varepsilon$. 

This work is organized as follows. In Sec. \ref{sec:theory}, we outline the fundamental theory and definition of the deflection angle. In Sec. \ref{sec:pertmeth} we present the main procedure of the perturbative method. The deflection angle for general SSS spacetime will be presented in this section. Sec. \ref{sec:applications} applies the deflection angle formula to the RN, charged Horndeski and charged Galileon spacetimes. The deflection is shown to be extremely accurate by comparison with numerical integration result. We conclude the paper with a discussion of our result in Sec. \ref{sec:conclusion}. Throughout this paper, we use the natural unit $G=c=\hbar=1$.

\section{Preliminary and the deflection angle}
	\label{sec:theory}

We consider a lens that is described by the general SSS metric
\begin{align}
\label{eq:spacetime}
\dd s^{2}
=
-A(r)\,\dd t^{2}
+B(r)\,\dd r^{2}
+C(r)
\left(
\dd\theta^{2}
+\sin^{2}\theta\,\dd\phi^{2}
\right),
\end{align}
where $(t,r,\theta,\phi)$ are the coordinates and $A(r)$, $B(r)$ and $C(r)$ are the metric functions. Although we can always choose $C(r)=r^2$, for now we will keep the general form of $C(r)$. For simplicity, in this work we will only consider the SSS that are asymptotically Minkowski and therefore $A(r)$ and $B(r)$ approach one asymptotically. 

We assume that the lens is surrounded by a homogeneous plasma. The refraction index of the plasma is $n=n(\omega)$, where $\omega$ denotes the frequency of the photon measured by a local observer at rest with respect to the plasma. Therefore,its frequency satisfies
\begin{align}
     \omega=-p_{\mu} v^{\mu} \label{eq;hbaromega},
\end{align}
where $v^\mu$ and $p_\mu$ are the four-velocity of the plasma and four-momentum of the photon respectively. The medium equation which links the four-momentum of the photon and $n$ is \cite{synge1960relativity} 
\begin{align}\label{eq:refractive index}
n^{2}=1+\frac{p_{\alpha}p^{\alpha}}{(p_{\beta}v^{\beta})^{2}}.
\end{align}

The photon paths through the plasma in the SSS spacetime can be described by the Hamiltonian in the geometric optics limit. Applying the variational principle
\begin{align}
    \delta \left(\int p_{\mu}dx^{\mu}\right)=0,
\end{align}
we can obtain the photon Hamiltonian \cite{synge1960relativity}
\begin{align}\label{eq:Hamiltonian}
    H(x^{\alpha},p_{\alpha})=\frac{1}{2} [g^{\alpha\beta}p_{\alpha}p_{\beta}-(n^{2}-1)(p_{\gamma}v^{\gamma})^2]=0,
\end{align}
where $n^2$ was given by Eq. \eqref{eq:refractive index}. Without loss of any generality, we will consider the photon path in the equatorial plane and therefore set $\theta=\pi/2$ in the metric \eqref{eq:spacetime} and correspondingly $p_\theta=0$ in the Hamiltonian.

For plasma of electrons and ions with fixed density $N$, its refractive index has the form of
\begin{align}\label{eq:refractive}
    n^{2}=1- \frac{\omega_{e}^{2}}{\omega^{2}(r)},\quad \omega_{e}^{2}=\frac{4\pi e^{2}N}{m},
\end{align}
where $\omega_{e}$ is the electron plasma frequency, $e$ is the charge of the electron and $m$ is the mass of the electron. Substituting Eq. \eqref{eq;hbaromega} and Eq. \eqref{eq:refractive} into Eq. \eqref{eq:Hamiltonian}, the Hamiltonian simplifies to
\begin{align}\label{eq:new H}
    H(x^{\alpha},p_{\alpha})=\frac{1}{2}(g^{\alpha\beta}p_{\alpha}p_{\beta}+\omega_{e}^2)=0.
\end{align}
The light-ray trajectories are then governed by Hamiltonian canonical equations
\begin{subequations}
\begin{align}    \label{eq:drodk}
\frac{\dd r}{\dd \lambda}=&\frac{ \partial H}{ \partial p_{r}}=\frac{p_{r}}{B(r)},\\
\label{eq:dphiodk}
\frac{\dd \phi}{\dd \lambda}=&\frac{ \partial H}{ \partial p_{\phi} }=\frac{p_{\phi}}{C(r)},\\
p_{t}=& -E, \label{eq:ptine}
\\  p_{\phi}=& L, \label{eq:pphiinl}
\end{align}
\end{subequations}
where $\lambda$ is the affine parameter along the photon trajectory.
Clearly, in the last two equations since the Hamiltonian \eqref{eq:Hamiltonian} does not depend on $t$ or $\phi$ explicitly, there exist conserved energy $E$ and angular momentum $L$ of the photon.

Denoting the asymptotic photon frequency as $\omega_\infty$, and the asymptotic refractive index as $n_\infty$, 
then the $n_\infty$ has a nontrivial relation with $\omega_e$ and $\omega_\infty$.
Under the assumption of a homogeneous plasma, it is clear that 
\begin{align}
    v^{t}=\sqrt{-g^{tt}},\quad v^{i}=0~(i=1,2,3).
\end{align}
Substituting these into Eq. \eqref{eq;hbaromega}, the local photon energy becomes 
\begin{align} 
 \omega=-p_{t}\sqrt{-g^{tt}}. \label{eq:homegasim}
\end{align}
Using Eq. \eqref{eq:ptine} and the metric function in Eq. \eqref{eq:spacetime}, we then obtain the 
effective redshift relation
\begin{align}
    \omega(r)=\frac{\omega_{\infty}}{A^{1/2}(r)}.
\end{align}
Substituting this into Eq. \eqref{eq:refractive}, we have
\begin{align}
     n_{\infty}=\lim_{ r \to \infty } \sqrt{1-\frac{\omega_{e}^{2}}{\omega^{2}(r)}}=\sqrt{1-\frac{\omega_{e}^{2}}{\omega_{\infty}^{2}}}.\label{eq:ninfdef}
\end{align}

The energy $E$ and angular momentum $L$ now can be linked to $\omega_\infty$ and $n_\infty$ using the relation
\begin{align}
E=\omega_{\infty},\quad L=b p_\infty=bn_{\infty}\omega_{\infty} , \label{eq:elinothers}
\end{align}
where $b$ is the impact parameter of the photon trajectory and $p_\infty$ is the asymptotic three-momentum of the photon. In the last step of Eq. \eqref{eq:elinothers}, we have used the relation $p_\infty=n_\infty\omega_\infty$
solved from the
dispersion relation 
\begin{align}
\omega_\infty^2=\omega_e^2+p_\infty^2
\end{align} and Eq. \eqref{eq:ninfdef}.
It is important to note that for the homogeneous plasma case as we are considering, the photon's asymptotic group velocity is not one but $n_\infty$.

Using the Hamiltonian \eqref{eq:new H} and metric \eqref{eq:spacetime}, we can express $p_r$ in terms of $p_t$ and $p_\phi$ as
\begin{align}
p_{r}=&\pm p_{t}\sqrt{ \frac{B(r)}{A(r)}-\frac{B(r)}{C(r)}\frac{p_{\phi^{2}}}{p_{t}^{2}}-B(r) \frac{\omega_{e}^{2}}{p_{t}^{2}} }\nonumber\\
    =&\mp \omega_{\infty}\sqrt{ \frac{B(r)}{A(r)}- \frac{B(r)}{C(r)}n_{\infty}^{2}b ^{2}-B(r) \frac{\omega_{e}^{2}}{\omega_{\infty}^{2}} },\label{eq:prsubed}
\end{align}
where in the second step we substituted Eqs. \eqref{eq:ptine} and \eqref{eq:pphiinl} and used Eq. \eqref{eq:elinothers}.
Then combining Eqs. \eqref{eq:drodk} and \eqref{eq:dphiodk} to eliminate the affine parameter $\lambda$ and replacing $p_\phi$ by Eq. \eqref{eq:pphiinl} and further by Eq. \eqref{eq:elinothers}, and $p_{r}$ by Eq. \eqref{eq:prsubed}, we can obtain
\begin{align}\label{eq:weifen}
    \frac{\dd \phi}{\dd r}=& \frac{p_{\phi}}{p_{r}} \frac{B(r)}{C(r)}\nonumber\\
=&\mp \frac{\sqrt{ A(r)B(r) }}{C(r)} \frac{n_{\infty}b}{\sqrt{\displaystyle  1-\frac{A(r)}{C(r)} n_{\infty}^{2} b^{2} -A(r) \frac{\omega_{e}^{2}}{\omega_{\infty}^{2}} }}.
\end{align}
The change of the angular coordinate $\Delta \phi$ from a source at radius $r_{i}$ to a detector at radius $r_{f}$ can be computed as
\begin{align}\label{eq:deflection angle}
    \Delta\phi=\sum_{j=i,f}\int_{r_{0}}^{r_{j}} \frac{\dd \phi}{\dd r} \dd r, 
\end{align}
where $\dd \phi/\dd r$ is given by Eq. \eqref{eq:weifen} and $r_{0}$ is the perihelion radius. It is defined by $\frac{\dd r}{\dd\lambda}\big|_{r=r_{0}}=0$, which after using Eqs. \eqref{eq:drodk} and \eqref{eq:prsubed} yields a relation between the impact parameter $b$ and the perihelion radius $r_0$ as
\begin{align}
    b=\frac{1}{n_{\infty}}\sqrt{ \frac{C(r_{0})}{A(r_{0})} -C(r_{0}) \frac{\omega_{e}^{2}}{\omega_{\infty}^{2}} } . 
    \label{eq:br0rel}
\end{align}

\section{Perturbative method}
\label{sec:pertmeth}

The integral in Eq. \eqref{eq:deflection angle} usually cannot be directly carried out for general metric functions and therefore necessitates some kind of approximation. In this section, we will present a perturbative method that was initiated in Ref. \cite{Huang:2020trl,Jia:2020xbc,Liu:2022hbp}. We will show that the method can be directly generalized to the computation of the deflection angle within a plasma. 

We first make a notation simplification to Eq. \eqref{eq:weifen} by introducing the frequency ratio $\varepsilon$ between the electron plasma frequency and far-side photon frequency 
\begin{align} \varepsilon^{2}=\frac{\omega_{e}^{2}}{\omega_{\infty}^{2}}=\frac{4\pi e^2N}{m\omega_\infty^2}, \label{eq:epsdef}
\end{align} 
where in the last step Eq. \eqref{eq:refractive} is used. 
We will also assume that the sender and receiver of the photon are both very far from the center so that the limit $r_{i}=r_f=\infty$ in Eq. \eqref{eq:deflection angle} can be taken. Then the angular change $\Delta\phi$  becomes
\begin{align}
    \Delta\phi=2\int_{r_{0}}^{\infty} \frac{\sqrt{ A(r)B(r) }}{C(r)} \frac{n_{\infty}b \dd r}{\sqrt{ 1-\frac{A(r)}{C(r)} n_{\infty}^{2} b^{2} -A(r) \varepsilon^2 }}, 
    \label{eq:dphisim}
\end{align}
where $n_\infty$, which was defined in Eq. \eqref{eq:ninfdef}, now can also be expressed through $\varepsilon$ as
\begin{align}
    n_\infty=\sqrt{1-\varepsilon^2}. \label{eq:ninfineps}
\end{align}

To eventually carry out the $r$ integral and obtain the series result of the deflection angle in the desirable form, the crucial next step is then to introduce the following changes of variables from $r$ to $x$ which are linked by 
\begin{align}
    r=\frac{1}{x},
\end{align}
and then from $x$ to $u$ linked by
\begin{align}\label{eq:var_substitution}
    u^{2}(x)=\frac{A(1/x)}{C(1/x)} n_{\infty}^{2} b^{2}+A(1/x) \varepsilon^2.
\end{align}
For later usage, we will denote the inverse function of $u^2(x)$ as $x(u^2)$.
The change of variables \eqref{eq:var_substitution} directly converts the large square root part in the denominator of Eq. \eqref{eq:dphisim} to $\sqrt{1-u^2}$, which is easily integrated. The integral limits in Eq.  \eqref{eq:dphisim} correspond to the limits $1/r_0$ and $0$ respectively for $x$. Using Eq. \eqref{eq:br0rel} and the asymptotic Minkowski condition that $A(\infty)=1$, they further correspond, respectively, to the following limit of $u$
\begin{align}
    u(x=1/r_0)=1,\quad u(x=1/\infty)=\varepsilon.
\end{align}
Therefore, superficially $\Delta\phi$ in Eq. \eqref{eq:dphisim} can be converted to an integral of $u$ from $\varepsilon$ to $1$, with all $r$ in the integrand now understood as a function of $u$. 

\begin{figure}[htp!]
    \centering
    \includegraphics[width=0.48\textwidth]{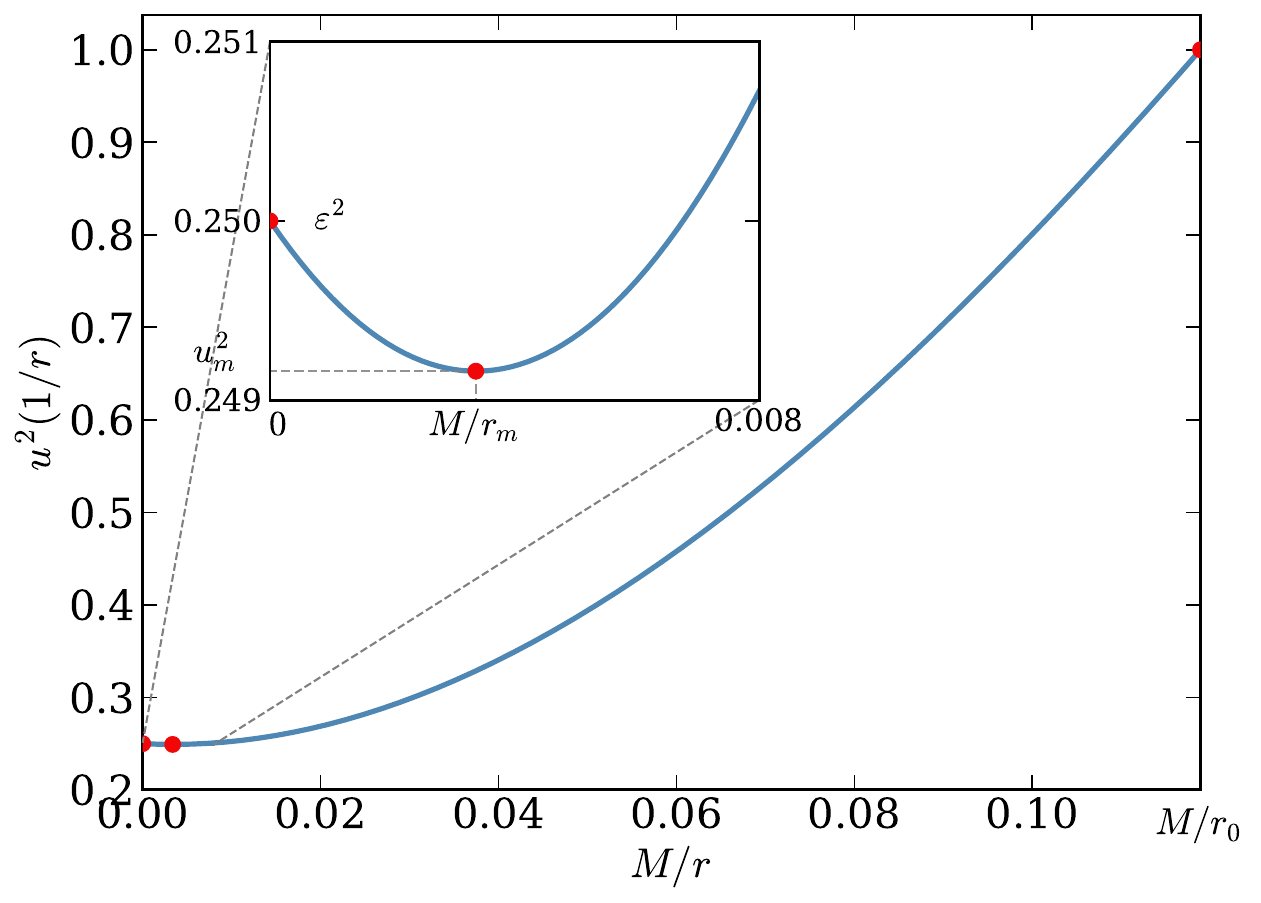}
    \caption{Function $u^2(1/r)$ defined in Eq. \eqref{eq:var_substitution}. We took the RN spacetime metric as an example when plotting. The parameters used are $Q=M/4,~\varepsilon=1/2,~b=10M$. The red dots correspond to the two boundaries of $1/r$ as well as the minimum point at $(1/r_\text{m},\,u_\text{m})$.}
    \label{fig:cov}
\end{figure}

However, it turns out that the function $u^2(x)$ in Eq. \eqref{eq:var_substitution} is not monotonic in the region of $r$ between $r_0$ and $\infty$ (see Fig. \ref{fig:cov}). Instead, it is found that there always exists an  $r_\text{m}\in(r_0,\,\infty)$, or equivalently an $x_\text{m}\equiv 1/r_\text{m}\in(0,\,1/r_0)$, such that $u^2(x)$ is monotonically decreasing for $x\in (0,\,1/r_\text{m})$ and monotonically increasing for $x\in[1/r_\text{m},\,1/r_0)$. We will denote the value of $u^2$ at $x_\text{m}=1/r_\text{m}$ as $u_\text{m}^2$, that is 
$ u_{\text{m}}^{2}=u^2(x_\text{m})$. This $x_m$ then can be solved using the condition 
\begin{align}
\left(u^2\right)^\prime(x)\big|_{x=x_\text{m}}=0.
\label{eq:xmcond}
\end{align}
Because of this non-monotonicity, when we substitute the inverse function $r=1/x(u^2)$ back into Eq. \eqref{eq:dphisim}, we will have to split the integral of $u$ into two parts. Denoting the inverse function of $u^2(x)$ for $x\in (0,\,1/r_\text{m})$ as $x_-(u^2)$ and the inverse function for $x\in (1/r_\text{m},\,1/r_0)$ as $x_+(u^2)$, finally the deflection angle \eqref{eq:dphisim} transforms into
\begin{align}
&\Delta\phi=2\int^{\varepsilon}_{u_{\text{m}}} \frac{n_{\infty}b}{\sqrt{ 1-u^2 }} \frac{\sqrt{ A(1/x_-(u^2))B(1/x_-(u^2)) }}{C(1/x_-(u^2))} J_-(u)\dd u\nonumber\\
&+2 \int_{1}^{u_{\text{m}}}  \frac{n_{\infty}b }{\sqrt{ 1-u^2 }}\frac{\sqrt{ A(1/x_+(u^2))B(1/x_+(u^2)) }}{C(1/x_+(u^2))}  J_+(u)\dd u,
\label{eq:dphitrans1}
\end{align}
where the two Jacobian factors associated with the change of variables are
\begin{align}
    J_\pm(u)=\frac{\dd r}{\dd x_\pm}\frac{\dd x_\pm}{\dd u}=-\frac{1}{x_\pm^2(u^2)}\frac{\dd x_\pm(u^2)}{\dd (u^2)}2u. 
    \label{eq:jacobdef}
\end{align}

Our next step is then to expand the integrands in Eq. \eqref{eq:dphitrans1} so that it becomes a much simpler series whose integration can be easily carried out. 
We first collect the factors containing $u$ in the integrands except the $1/\sqrt{1-u^2}$ factor and denote them as functions $h_\pm(u)$, i.e.,
\begin{align}
    h_\pm(u)=\frac{\sqrt{ A(1/x_\pm(u^2))B(1/x_\pm(u^2)) }}{C(1/x_\pm(u^2))}&  J_\pm(u). \label{eq:hdef}
\end{align}
Our plan is to expand these functions around $u=u_\text{m}$. To achieve this, clearly we need to have the metric functions $A,\,B,\,C$ expanded for $1/x_\pm$
first and then $x_\pm$ expanded around $u=u_\text{m}$. In this work, since we are seeking a deflection angle in the weak field limit, we can assume that the metric functions take the following form of asymptotic expansions
\begin{align}\label{eq:metric expansion}
    A(r)=1+\sum^{\infty}_{j=1} \frac{a_j}{r^j},\; B(r)=1+\sum^{\infty}_{j=1} \frac{b_j}{r^j},\; C(r)=r^2,
\end{align}
where $a_j$ and $b_j$ are the expansion coefficients and here and henceforth, we fix $C(r)=r^2$. We note that the coefficient $a_1$ usually can be identified as $-2M$ where $M$ is the ADM mass of the spacetime. We remind that once the spacetime is known, these expansions can be easily obtained, and the spacetime parameters are all encoded in the expansion coefficients. 

For the expansion of $x_\pm(u^2)$, due to the fact that $u_m$ is a branch point of the function, the generalized Lagrange inversion theorem dictates that they must take the following form
\begin{align}
    x_\pm(u^2)=\sum_{j=0}^\infty x_{j}\left[\pm(u^2-u_\text{m}^2)^{1/2}\right]^j. \label{eq:xexp}
\end{align}
The coefficients $x_{j}$ here can be obtained by substituting this, as well as the expansions \eqref{eq:metric expansion}, into the defining relation \eqref{eq:var_substitution}
and then using the undetermined coefficient method. It is found that all the $x_{j}~(j=1,\,\cdots)$ can be expressed iteratively as polynomial functions of coefficients of lower order, i.e., $\{x_{0},\,\cdots,\,x_{j-1}\}$, as well as the metric functions'  asymptotic expansion coefficients. However, their forms are quite lengthy and therefore we will not present them here. 
The leading coefficient $x_0$ in Eq. \eqref{eq:xexp}, by definition, is where the function $u^2(x)$ reaches its minimum and therefore $x_0\equiv x_\text{m}$. Its value should satisfy the requirement Eq. \eqref{eq:xmcond}. This condition, however, is also a series equation and therefore we will also solve it perturbatively. To this purpose, one can use the following ansatz for the series of $x_m$  
\begin{align}
    x_\text{m}=\sum_{j=1}^\infty x_{\text{m},j} b^{-2j}, \label{eq:xmexp}
\end{align}
and solve the coefficients $x_{\text{m},j}$ by expanding Eq. \eqref{eq:xmcond} in large $b$. The result for the first few terms is found to be
\begin{subequations}
    \begin{align}
    x_{\text{m},1}=&-\frac{a_{1}\varepsilon^{2}}{2n_{\infty}^{2}}, \\
    x_{\text{m},2}=& \frac{a_{1}(3a_{1}^{2}-4a_{2})\varepsilon^{4}}{8n_{\infty}^{4}},\\
    x_{\text{m},3}=& \frac{a_{1}(9a_{1}^{4}-22a_{1}^{2}a_{2}+8a_{2}^{2}+6a_{1}a_{3})\varepsilon^{6}}{16n_{\infty}^{6}}.
\end{align}
\end{subequations}
It is clear that they are also expressed as polynomial functions of the metric expansion coefficients, as well as the parameters $b,\, n_\infty$ and $\varepsilon$ appearing in Eq. \eqref{eq:var_substitution}.
For later usage in the limits of the integrals in Eq. \eqref{eq:dphitrans1}, we can now solve for $u_\text{m}$ by substituting the obtained $x_\text{m}$ in Eq. \eqref{eq:xmexp} into Eq. \eqref{eq:var_substitution}. It is readily found that 
\begin{align}
    u_\text{m}=\sum_{j=0}^\infty u_{\text{m},j} b^{-2j}, \label{eq:umexp}
\end{align}
where
\begin{subequations}
    \begin{align}
    u_{\text{m},0}=&\varepsilon^{2}, \\
    u_{\text{m},1}=& -\frac{a_{1}^{2}\varepsilon^{4}}{4n_{\infty}^{2}},\\
    u_{\text{m},2}=& \frac{a_{1}^{2}(a_{1}^{2}-2a_{2})\varepsilon^{6}}{8n_{\infty}^{4}}.
\end{align}
\end{subequations}

Now with $x(u^2)$ known in Eq. \eqref{eq:xexp}, we can substitute it back into the integrand factors $h_\pm(u)$ in Eq. \eqref{eq:hdef}. They are found to also take a power series of $(u^2-u_\text{m}^2)$ in the form
\begin{align}
h_\pm(u^2)=\sum_{j=-1}^{\infty}h_ju\left[\pm(u^{2}-u_\text{m}^2)^{1/2}\right]^j, \label{eq:hexp}
\end{align}
where the first $u^1$ factor is inherited from the Jacobian factors \eqref{eq:jacobdef}. The coefficients $h_j$ are, unfortunately, also too lengthy to present here but with a computer algebra, they are straightforward to obtain. It is found that they are 
also functions of the asymptotic expansion coefficients of the metric functions, and functions of $x_\text{m}$ and other parameters. 
Substituting Eq. \eqref{eq:hexp} into the deflection angle \eqref{eq:dphitrans1}, $\Delta\phi$ is expressed as
\begin{align}
    \Delta\phi=2n_\infty b \sum_{j=-1}^\infty h_j(I_{+j}+I_{-j}), \label{eq:dphitrans2}
\end{align}
where for $j=-1,\,0,\,\cdots$
\begin{subequations}
    \begin{align}
        I_{+j}=(+1)^j \int_{1}^{u_{\text{m}}}\frac{u(u^{2}-u_{\text{m}}^2)^{j/2}}{\sqrt{ 1-u^{2} }}\dd u,\\
       I_{-j}=(-1)^j \int_{u_{\text{m}}}^\varepsilon\frac{u(u^{2}-u_{\text{m}}^2)^{j/2}}{\sqrt{ 1-u^{2} }}\dd u.
    \end{align}
\end{subequations}
The integrals in $I_{\pm j}$ are elementary and the results are mostly polynomials of $u_\text{m}$
with a few terms involving the $\arctan$ function. Now since $u_\text{m}$ as given in Eq. \eqref{eq:umexp} is already a series of $1/b$, it only makes sense to also expand $\Delta\phi$ in Eq. \eqref{eq:dphitrans2} for small $1/b$. After reorganizing into a series of $1/b$, it becomes
\begin{align}
\Delta \phi= \sum_{j=0}^{\infty} \frac{d_j}{b^j}\equiv\sum_{j=0}^{\infty}\Delta \phi_j,  \label{eq:dphif1}
\end{align}
where in the last step we defined the notation $\Delta\phi_j$ for later easier reference. The first few orders of $d_j$ are given by the following with $n_\infty$ in Eq. \eqref{eq:ninfineps}
\begin{subequations}
\label{eq:dphiinb}
    \begin{align}
        d_0=&\pi,\\
    d_1=&-\frac{a_{1}}{n_{\infty}^{2}}+b_{1},\\
        d_2=&\frac{\pi}{4n_{\infty}^{2}}(2a_{1}^{2}-2a_{2}-a_{1}b_{1})+\frac{\pi}{16}(-b_{1}^{2}+4b_{2}),\\
           d_{3}=&\frac{a_{1}^{3}}{12n_{\infty}^{6}}+\frac{a_{1}}{4n_{\infty}^{4}}(-4a_{1}^{2}+4a_{2}+a_{1}b_{1})+\frac{1}{4n_{\infty}^{2}}(-8a_{1}^{3}\nonumber\\
           &+16a_{1}a_{2}-8a_{3}+4a_{1}^{2}b_{1}+a_{1}b_{1}^{2}+4a_{1}b_{2}-4a_{2}b_{1})\nonumber\\
           &+\frac{1}{12}(32a_{3}+b_{1}^{3}-4b_{1}b_{2}+8b_{3}),\\
            d_{4}=&\frac{3\pi}{64n_{\infty}^{4}}(24a_{1}^{4}-48a_{1}^{2}a_{2}+16a_{1}a_{3}+8a_{2}^{2}-8a_{1}^{3}b_{1}\nonumber\\
            &-a_{1}^{2}b_{1}^{2}+8a_{1}a_{2}b_{1}+4a_{1}^{2}b_{2})+\frac{3\pi}{64n_{\infty}^{2}}(16a_{1}^{4}\nonumber\\
            &-48a_{1}^{2}a_{2}+16a_{2}^{2}+32a_{1}a_{3}-16a_{4}-8a_{1}^{3}b_{1}\nonumber\\
            &-2a_{1}^{2}b_{1}^{2}-a_{1}b_{1}^{3}+4a_{1}b_{1}b_{2}+8a_{1}^{2}b_{2}-8a_{1}b_{3}\nonumber\\
            &+16a_{1}a_{2}b_{1}+2a_{2}b_{1}^{2}-8a_{2}b_{2}-8a_{3}b_{1})\nonumber\\
            &+\frac{3\pi}{1024}(-960a_{1}a_{3}+320a_{4}+160a_{3}b_{1}\nonumber\\
            &-5b_{1}^{4}+24b_{1}^{2}b_{2}-32b_{1}b_{3}-16b_{2}^{2}+64b_{4}).
    \end{align}
\end{subequations}

In reality, the plasma density $N$ is usually small and therefore in this work we will assume that $\omega_e\ll \omega(r)$ and therefore $\varepsilon$ is a very small parameter. Many previous studies have worked in the limit $\varepsilon\to 0$. Therefore for reference and comparison purposes, we can also expand the deflection angle in  Eq. \eqref{eq:dphif1} into a series of $\varepsilon^2$ so that the result becomes a dual series of $1/b$ and $\varepsilon$
\begin{align}
\Delta \phi= \sum_{j,k=0}^{\infty} d_{j,k}\frac{\varepsilon^{2k}}{b^j}~\text{where}~ \varepsilon^2=\frac{4\pi e^2N}{m\omega_\infty^2}.  \label{eq:dphif2}
\end{align}
The coefficients $d_{j,K}$ can be easily worked out, with the first few ones up to $j+k=3$ given by
\begin{subequations}
\label{eq:dphiinbeps}    
\begin{align}
d_{0,0}=&\pi,\,d_{0,1}=d_{0,2}=d_{0,3}=0,\\
d_{1,0}=&-a_{1}+b_{1},\\
d_{1,1}=&-a_{1},\\
d_{1,2}=&-a_{1},\\
d_{2,0}=&\frac{\pi}{16}(8a_{1}^{2}-8a_{2}-4a_{1}b_{1}-b_{1}^{2}+4b_{2}),\\
d_{2,1}=&\frac{\pi (a_{1}^{2}-a_{2}-a_{1}b_{1})}{2},\\
d_{3,0}=&\frac{1}{12}(-35a_{1}^{3}+60a_{1}a_{2}+8a_{3}+15a_{1}^{2}b_{1}+3a_{1}b_{1}^{2}\nonumber\\
&-12a_{1}b_{2}-12a_{2}b_{1}+b_{1}^{3}-4b_{1}b_{2}+8b_{3}).
    \end{align}
\end{subequations}
Due to the space limit, here we only showed the series up to order $\varepsilon^4$. However, unlike the series in $1/b$, it is very easy to obtain higher-order results since the series in $\varepsilon$ is obtained from the expansion of Eq. \eqref{eq:dphiinb}.

To our best knowledge, Eqs. \eqref{eq:dphiinb} and \eqref{eq:dphiinbeps} are the first results for the deflection angle of light rays in the weak field limit in dilute homogeneous plasma in general SSS spacetime. We observe that in the no plasma limit ($n_\infty\to 1$ or $\varepsilon\to0$), these equations reduce to known deflection angles in pure SSS spacetime,  e.g. Eq. (4.8) and Eq. (B.1) of Ref. \cite{Huang:2020trl}.
For the effect of the plasma, a few remarks are worth mentioning here. The first is that, as $d_{0,k\geq 1}=0$, the plasma effect does not appear in the zero-th order of $M/b$ where $M$ stands for some kind of mass of the central object. This agrees with physical intuition that for a gravity field with zero $M$, the uniform plasma would not bend the photon trajectory either. Rather, the $\varepsilon^2$ term shows up from the $\mathcal{O}(M/b)^1$ order. Then the second point is that, since $\varepsilon^2<1$ but is independent of the kinetic variable $b$, the plasma effect will always be larger than the second order $\mathcal{O}(M/b)^2$ for large enough impact parameter. The third point is that since $d_{1,1}=-a_1=2M$, the plasma effect, in its leading order, always enlarges the deflection angle of the photon. 

\section{Applications}
\label{sec:applications}

In this section, we will apply the results derived for the general SSS spacetime in the last section, Eqs. \eqref{eq:dphiinb} and \eqref{eq:dphiinbeps}, to several representative spacetimes to obtain the deflection angles in them. We will compare these deflection angles with numerical integration results and show the accuracy of our perturbative formulas. Comparison will also be made with previous literature dealing with the same problem in the homogeneous plasma but only within specific spacetimes.
Deflection angles in more spacetimes will be given in the Appendix \ref{sec:app1}. 

\subsection{Deflection in RN spacetime}
\label{subsec:RN}

We first consider the deflection of light rays in RN spacetime described by the metric functions
\begin{align}
    A(r)=\frac{1}{B(r)}=1-\frac{2M}{r}+\frac{Q^{2}}{r^{2}},\label{eq:rnmetric}
\end{align}
where $M$ and $Q$ are the spacetime mass and electric charge, respectively. Results here will cover those in the Schwarzschild spacetime when we set $Q=0$. 

The metric functions \eqref{eq:rnmetric} can be easily expanded asymptotically according to the form \eqref{eq:metric expansion} to find the coefficients 
\begin{subequations}
\begin{align}
        &a_{1}=-2M, \quad  a_{2}=Q^{2},\quad  a_{n}=0 ~(n\geq 3 ),\\
&b_{1}=2M,\quad b_{2}=4M^{2}-Q^{2}   ,\quad b_{3}=8M^{3}-4MQ^{2},\nonumber\\
&b_{4}= 16M^{4}-12M^{2}Q^{2}+Q^{4},~\dots.
    \end{align}
\end{subequations}    
Substituting these coefficients into Eqs. \eqref{eq:dphif1}, we can obtain the deflection angle in the RN spacetime with a homogeneous plasma as a single series of $M/b$
\begin{align}
    &\Delta\phi_\text{RN}=  \pi+\frac{2M}{b}\left( 1+\frac{1}{n_{\infty}^{2}} \right)+ \frac{3\pi M^{2}}{4b ^{2}}\left[ 1+\frac{4}{n_{\infty}^{2}}\right.\nonumber\\
    &\left.-\frac{Q^{2}}{3M^{2}}\left( 1+\frac{2}{n_{\infty}^{2}} \right) \right]+\frac{10M^{3}}{3b^{3}}\left[ 1+\frac{9}{n_{\infty}^{2}}+\frac{3}{n_{\infty}^{4}}-\frac{1}{5n_{\infty}^{6}}\right.\nonumber\\
    &\left.-\frac{3Q^{2}}{5M^{2}}\left( 1+\frac{6}{n_{\infty}^{2}}+\frac{1}{n_{\infty}^{4}} \right) \right]\nonumber\\
    &+\frac{105\pi M^{4}}{64b^{4}}\left[ 1+\frac{16}{n_{\infty}^{2}}+\frac{16}{n_{\infty}^{4}}-\frac{6Q^{2}}{7M^{2}}\left( 1+\frac{12}{n_{\infty}^{2}}+\frac{8}{n_{\infty}^{4}} \right)\right.\nonumber\\
    &\left.+\frac{3Q^{4}}{35M^{4}}\left( 1+\frac{8}{n_{\infty}^{2}}+\frac{8}{3} \frac{1}{n_{\infty}^{4}} \right) \right]+\mathcal{O}\left(\frac{M^5}{b^5}\right).\label{eq:dphirn1}
\end{align}
Ref. \cite{Matsuno:2020kju} considered the deflection in a RN-type spacetime. Our result here reduces to its Eq. (26), which contains a result to the order $M^2/b^2$.
When setting $\varepsilon=0$, Eq. \eqref{eq:dphirn1} reduces to the deflection angle of null rays in pure RN spacetime, which agrees with Eq. (54) of Ref. \cite{Jia:2020xbc}.
In the small $\varepsilon$ limit, Eq. \eqref{eq:dphirn1}
can be written in a dual series form
\begin{align}
&\Delta\phi_{\text{RN}}=\pi+\frac{4M}{b}+\frac{3\pi(5M^{2}-Q^{2})}{4b ^{2}}\nonumber\\
&+\frac{16M(8M^{2}-3Q^{2})}{3b^{3}}+\frac{105\pi(33M^{4}-18M^{2}Q^{2}+Q^{4})}{64b^{4}}\nonumber\\
&+\varepsilon^{2}\left[ \frac{2M}{b}+\frac{\pi (6M^{2}-Q^{2})}{2b ^{2}}+\frac{16M(3M^{2}-Q^{2})}{b^{3}}\right.\nonumber\\
&\left.+\frac{15\pi (42M^{4}-21M^{2}Q^{2}+Q^{4})}{8b^{4}}\right]+\mathcal{O}\left( \varepsilon ^{4}, \frac{M^5}{b^{5}} \right).
\label{eq:dphirn2}
\end{align}
It is appropriate here to point out that the deflection angles in two other spacetimes with a homogeneous plasma can be obtained from Eq. \eqref{eq:dphirn1} and \eqref{eq:dphirn2} by mapping some parameters. For details, please see Eqs. \eqref{eq:EDM BH metric} to \eqref{eq:smres} in Appendix \ref{sec:app1}.

For completeness and future reference, here we also present the deflection angle in Schwarzschild spacetime with a homogeneous plasma obtainable by setting $Q=0$ in Eqs. \eqref{eq:dphirn1}
\begin{align}
    \Delta\phi_\text{S}=& \pi+\frac{2M}{b}\left( 1+\frac{1}{n_{\infty}^{2}} \right)+ \frac{3\pi M^{2}}{4b ^{2}}\left( 1+\frac{4}{n_{\infty}^{2}} \right)\nonumber\\
    &+\frac{10M^{3}}{3b^{3}}\left( 1+\frac{9}{n_{\infty}^{2}}+\frac{3}{n_{\infty}^{4}}-\frac{1}{5n_{\infty}^{6}} \right)\nonumber\\
    &+\frac{105\pi M^{4}}{64b^{4}}\left( 1+\frac{16}{n_{\infty}^{2}}+\frac{16}{n_{\infty}^{4}} \right)+\mathcal{O}\left(\frac{M^5}{b^5}\right) .\label{eq:dphisch1}
\end{align}
The low order version of this formula (to order $\mathcal{O}(M/b)^1$) was obtained in Eq. (32) of Ref. \cite{Crisnejo:2018uyn} and Eq. (38) of Ref. \cite{Bisnovatyi-Kogan:2010flt}. In the small $\varepsilon$ limit, Eq. \eqref{eq:dphisch1} becomes
    \begin{align}
\Delta\phi_{\text{S}}=&\pi+\frac{4M}{b}+ \frac{15\pi M^{2}}{4b^{2}}+ \frac{128M^{3}}{3b^{3}}+\frac{3465\pi M^{4}}{64b^{4}}\nonumber\\
&+\varepsilon^{2}\left( \frac{2M}{b}+\frac{3\pi M^{2}}{b ^{2}}+\frac{48M^{3}}{b^{3}}+\frac{315\pi M^{4}}{4b^{4}} \right)\nonumber\\
&+\mathcal{O}\left( \varepsilon ^{4}, \frac{M^5}{b^{5}} \right). \label{eq:dphisch2}
\end{align}

\begin{figure}[htp!]
\includegraphics[width=0.48\textwidth]{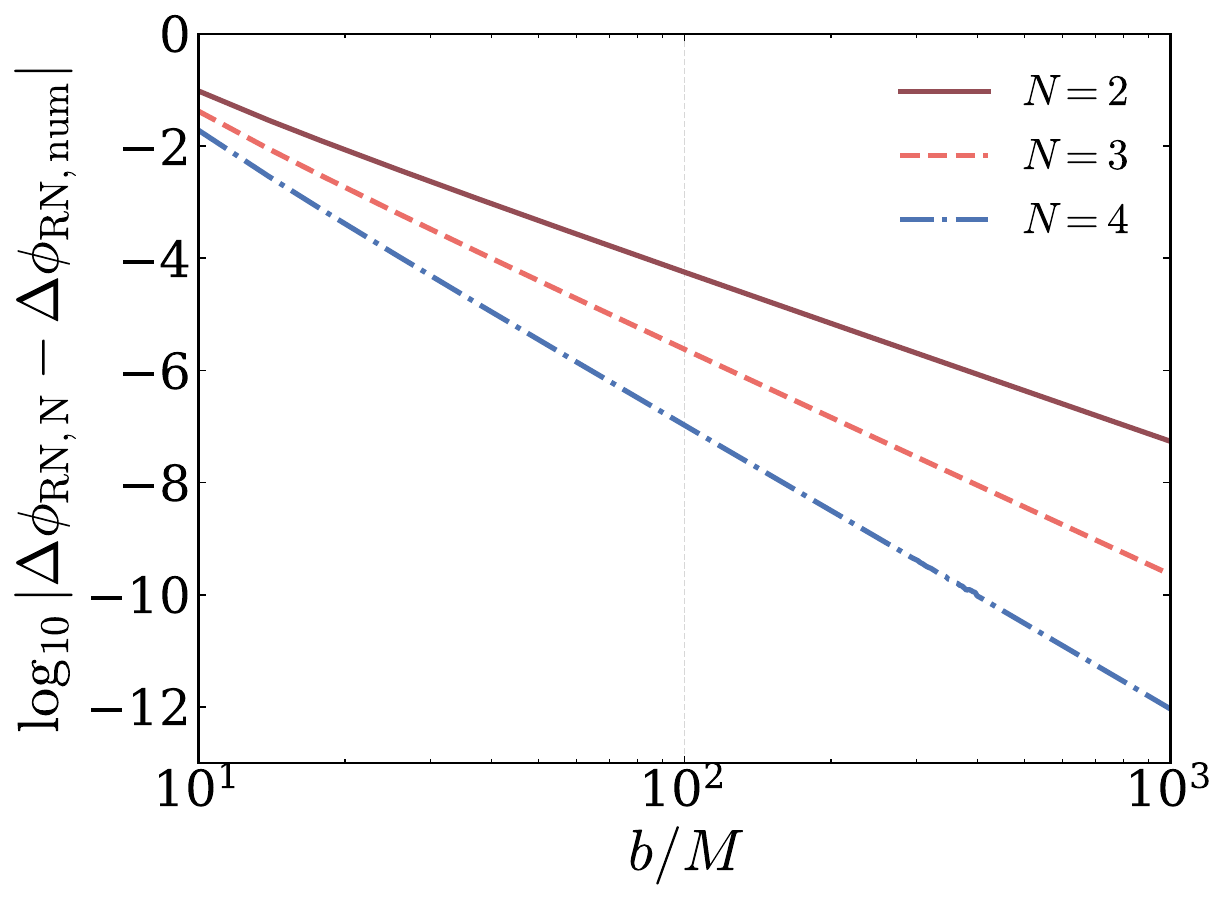}
\caption{\label{fig:rn1} Difference between deflection angle $\Delta\phi_{\text{RN},N}$ in Eq. \eqref{eq:dphirn1} but truncated to orders $N=2,\,3,\,4$ and the deflection angle $\Delta\phi_{\text{RN,num}}$ obtained by numerical integration. 
Parameter values $Q=M/2, \varepsilon=1/2$ were used.
}
\end{figure}

To further check the correctness of the deflection angles \eqref{eq:dphirn1} and \eqref{eq:dphirn2}, next we compare them with the result obtained using numerical integration of the original definition \eqref{eq:deflection angle}. We will also study the effect of various parameters, particularly the plasma density $\varepsilon$, on the deflection angles by plotting their variations with these parameters. 
In Fig. \ref{fig:rn1}, we plot the difference between the deflection \eqref{eq:dphirn1} truncated to orders $N=2$, 3 and 4, respectively, and the result obtained by numerically integrating Eq. \eqref{eq:deflection angle} for the RN metric. It is seen that as $N$ increases, the accuracy of the series result increases exponentially. Moreover, as a series of $M/b$, the result is clearly more accurate for larger $b$. To the truncation order $N=4$, we see that the difference between the series and numerical result has already reached a level of $10^{-12}$ for $b/M=10^3$, which shows the high reliability of these perturbative results.

\begin{figure}[htp!]
\includegraphics[width=0.44\textwidth]
{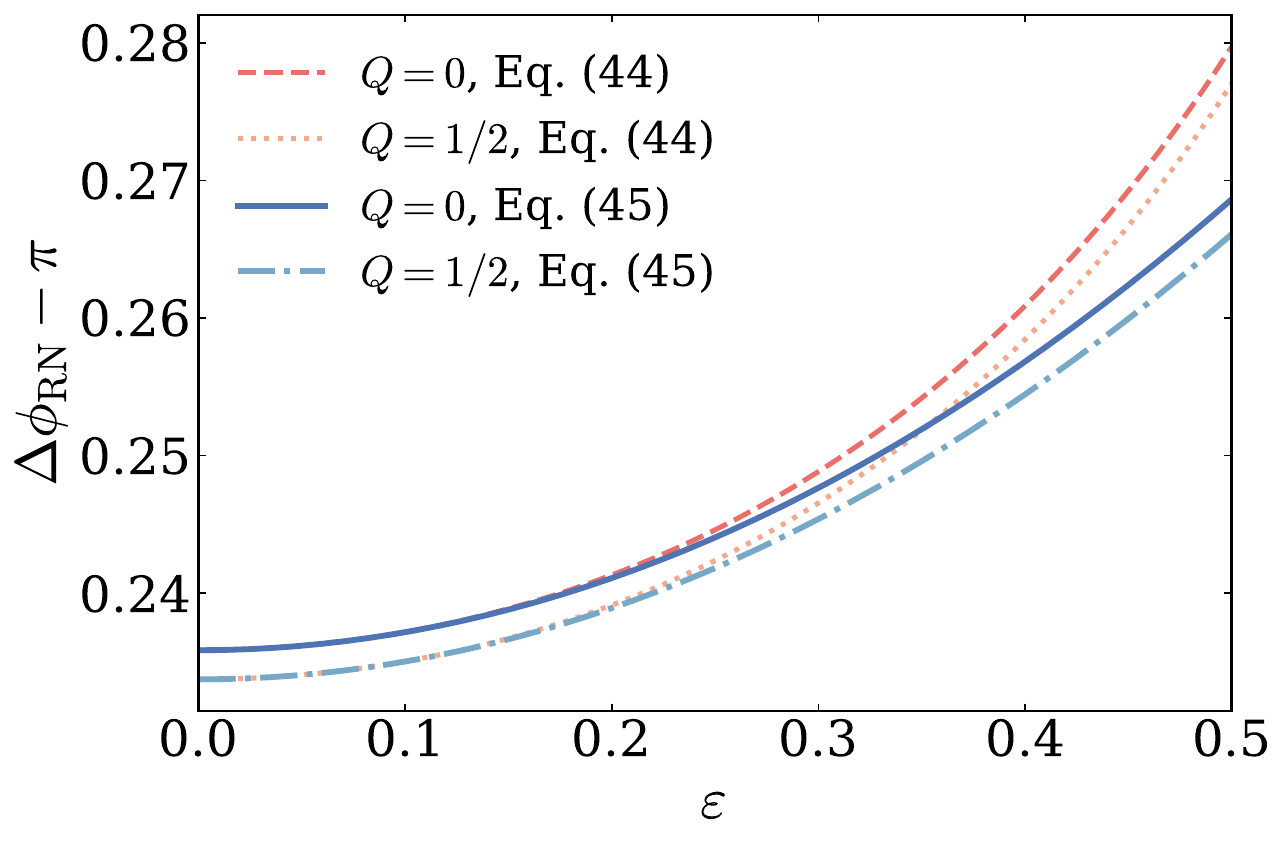}
\caption{\label{fig:rn2} Deflection angle \eqref{eq:dphirn1} and \eqref{eq:dphirn2} as a function of $\varepsilon$ for different $Q$'s. 
$b=20M$ is used in plotting the figure.
}
\end{figure}

In Fig. \ref{fig:rn2}, we plot the effect of $\varepsilon$ for a few $Q/M$ using both Eqs. \eqref{eq:dphirn1} and \eqref{eq:dphirn2}. It is seen that the deflections in all cases increase monotonically as $\varepsilon$ increases. This is consistent with the dependence on $\varepsilon^2$ in Eq. \eqref{eq:dphirn2}. Note that the curves plotted using Eq. \eqref{eq:dphirn1}, i.e. the full dependence on $\varepsilon$, grow slightly faster than the small $\varepsilon$ expanded formula Eq. \eqref{eq:dphirn2}. In addition, in these plots, we chose relatively small $b/M$ and large $\varepsilon$ to make the plasma effect distinguishable in the figure. In relativistic gravitational lensing, $b/M$ will be much larger and $\varepsilon$ is also much smaller  \cite{Ginburg1970}. The enhancement of $\Delta\phi$ by the plasma indicates that the apparent angles of the gravitational lensing images will also be larger than if there were no plasma. 

If we compare the effect of the charge $Q$ and plasma $\varepsilon$, we see from Eq. \eqref{eq:dphirn2} that their leading order contributions have different signs and their relative size depends on $\displaystyle \frac{Q^2}{M^2}\frac{M}{b}$ and $\displaystyle \varepsilon^2=\frac{4\pi e^2 N}{m\omega_\infty^2}$ respectively. This means that the charge effect and the plasma effect naturally compete in their influence on the deflection angle and their sizes might vary according to a few parameters. The question of under what circumstances that the plasma effect might be stronger/weaker than the central charge effect, and what the consequences are for gravitational lensing, will have to be studied in future more pragmatic works.

\subsection{Deflection in charged Horndeski spacetime}
\label{subsec:Charged Horndeski}

The metric functions of charged Horndeski spacetime are given by \cite{Feng:2015wvb}
\begin{subequations}
    \begin{align}
        &A(r)=1-\frac{2M}{r}+\frac{Q^{2}}{4r^{2}}-\frac{Q^{4}}{192r^{4}},\\
        &B(r)=\frac{1}{A(r)}\left( 1-\frac{Q^{2}}{8r^{2}} \right)^{2},
    \end{align}
\end{subequations}
where $M$ and $Q$ are the mass and charge of the spacetime, respectively. For the existence of the event horizon, $Q$ has to satisfy $|Q|< \frac{3\sqrt{ 2 }}{4}M$.
Expanding the above functions asymptotically, their expansion coefficients in the form of Eq. \eqref{eq:metric expansion} are
\begin{subequations}
\label{eq:chexp}
    \begin{align}
            a_{1}&=-2M,\quad a_{2}=\frac{Q^2}{4},\quad a_{3}=0 ,\nonumber\\
            a_{4}&=-\frac{Q^4}{192},\quad a_{n}=0,\quad(n\geq 5)\\
            b_{1}&=2M,\quad b_{2}=4M^{2}-\frac{Q^{2}}{2},\quad b_{3}=8M^{3}-\frac{3M Q^{2}}{2},\nonumber\\
b_{4}&=\frac{768M^{4}-192M^2Q^2+7Q^4}{48},\quad \dots
    \end{align}
\end{subequations}

Substituting these coefficients into Eqs. \eqref{eq:dphif1} and \eqref{eq:dphif2}, the deflection angles in the charged Horndeski spacetime with a homogeneous plasma are obtained as 
\begin{align}
&\Delta\phi_{\text{CH}}=\pi+\frac{2M}{b}\left( 1+\frac{1}{n_{\infty}^{2}} \right)+\frac{3\pi M^{2}}{4b ^{2}}\left[ 1+\frac{4}{n_{\infty}^{2}}\right.\nonumber\\
&\left.-\frac{Q^{2}}{6M^{2}}\left( 1+\frac{1}{n_{\infty}^{2}} \right) \right]+ \frac{10M^{3}}{3b^{3}}\left[ 1+\frac{9}{n_{\infty}^{2}}+\frac{3}{n_{\infty}^{4}}-\frac{1}{5n_{\infty}^{6}}\right.\nonumber\\
&\left.-\frac{Q^{2}}{5M^{2}}\left( 1+\frac{21}{4n_{\infty}^{2}}+\frac{3}{4n_{\infty}^{4}} \right) \right]+\frac{105\pi M^{4}}{64b^{4}}\nonumber\\
&\times\left[ 1+\frac{16}{n_{\infty}^{2}}+\frac{16}{n_{\infty}^{4}}-\frac{9Q^{2}}{35M^{2}}\left( 1+\frac{34}{3n_{\infty}^{2}}+\frac{64}{9n_{\infty}^{4}} \right)\right.\nonumber\\
&\left.+\frac{11Q^{4}}{1680M^{4}}\left( 1+\frac{100}{11n_{\infty}^{2}}+\frac{24}{11n_{\infty}^{4}} \right) \right]+\mathcal{O}\left(\frac{M^5}{b^5}\right) ,\label{eq:dphich1}
\end{align}
whose small $\varepsilon$ expansion is
\begin{align}
&\Delta\phi_{\text{CH}}=\pi+\frac{4M}{b}+\frac{\pi(15 M^{2}-Q^{2})}{4b ^{2}}+\frac{2M(64M^{2}-7Q^{2})}{3b^{3}}\nonumber\\
&+\frac{15\pi( 3696M^{4}-35M^{2}Q^{2}+9Q^{4})}{1024b^{4}}\nonumber\\
&+\varepsilon^{2}\left[ \frac{2M}{b}+\frac{\pi(24 M^{2}-Q^{2})}{8b ^{2}}+\frac{3M(32M^{2}-3Q^{2})}{2b^{3}}\right.\nonumber\\
&\left.+\frac{\pi(20160M^{4}-2760M^{2}Q^{2}+37Q^{4})}{256b^{4}} \right]+\mathcal{O}\left( \varepsilon^{4}, \frac{M^{5}}{b^{5}} \right). 
\label{eq:dphich2}
\end{align}
When $Q=0$, clearly Eqs. \eqref{eq:dphich1} and \eqref{eq:dphich2} reduce to the Schwarzschild result in Eqs. \eqref{eq:dphisch1} and \eqref{eq:dphisch2} respectively. 
When $\varepsilon=0$, Eq. \eqref{eq:dphich1} to the first three orders reduces to the known deflection in pure charged Horndeski spacetime \cite{Wang:2019cuf}. 

\begin{figure}[htbp]
\includegraphics[width=0.44\textwidth]
{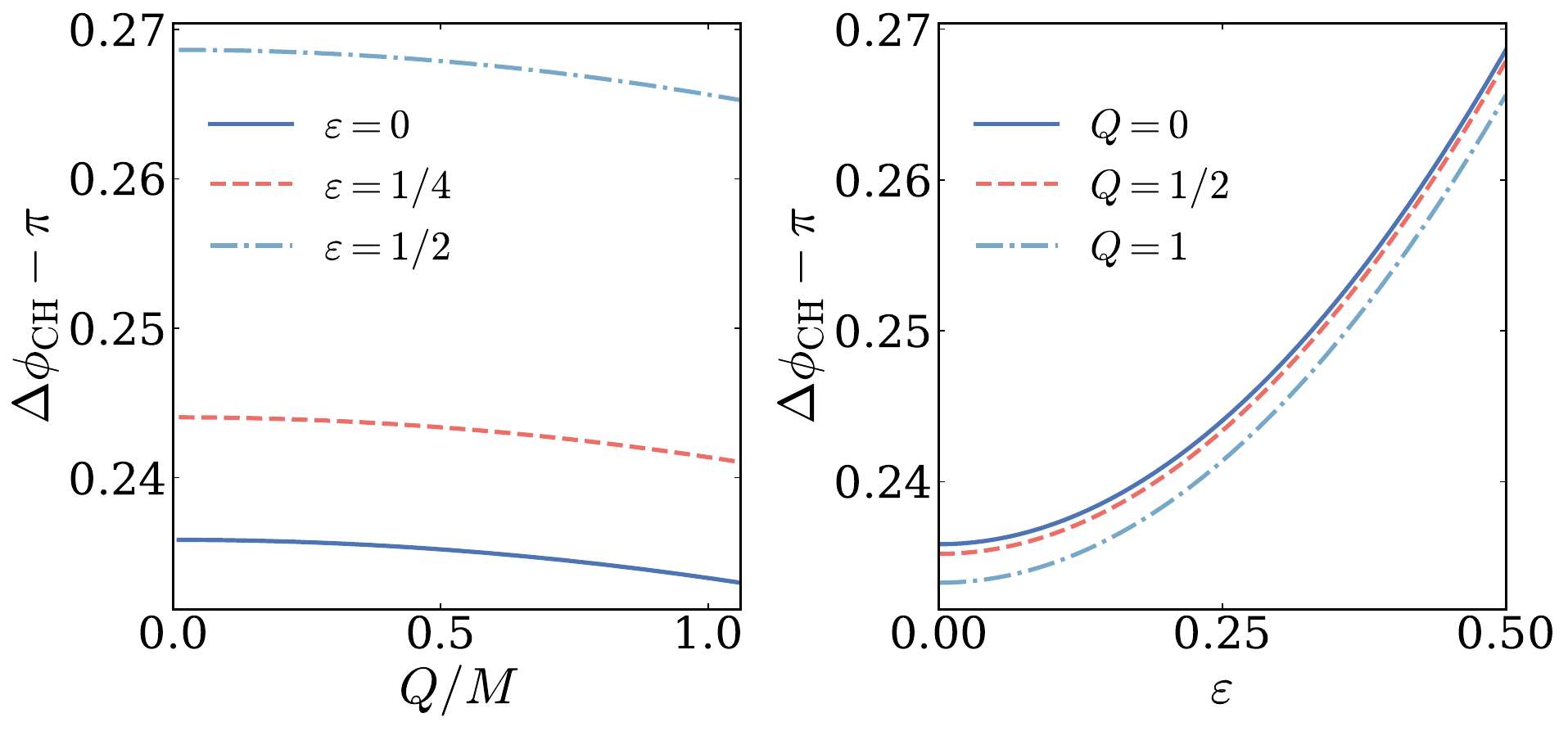}
\caption{\label{fig:ch1} Deflection angle \eqref{eq:dphich1} as a function of $Q$ (left) and $\varepsilon$ (right). 
$b=20M$ is used in plotting this figure.
}
\end{figure}

In Fig. \ref{fig:ch1} we plot the dependence of the deflection angle on the charge $Q$ and plasma parameter $\varepsilon$. It is seen from the left plot that as $Q$ increases, the deflection angle decreases for all $\varepsilon$. This is consistent with the effect of electric charge on the deflection angle in RN spacetime \cite{Pang:2018jpm}, although the charge here does not affect the spacetime structure exactly the same way as in the RN case. For the effect of the plasma, as shown in the right plot, again we see that $\Delta\phi$ increases monotonically as $\varepsilon$ increases for all $Q$. Comparing the left and right plots, we also observe that for the chosen parameter settings in the figure, the effect of the frequency ratio $\varepsilon$ is stronger than that of $Q$.
\subsection{Deflection in charged Galileon BH spacetime}
\label{subsec:Charged Galileon}

As the last example, we consider the deflection in the charged Galileon BH spacetime. The metric of the spacetime is given by \cite{Babichev:2015rva}
\begin{subequations}
    \begin{align}
        &A(r)=1-\frac{2M}{r}+\frac{\Gamma}{r ^{2}},\\
        &B(r)=\frac{1}{A(r)}\left( 1+\frac{\Gamma}{r^{2}} \right)^{-1},
    \end{align}
\end{subequations}
where 
\begin{align}
    \Gamma= \frac{Q^{2}+P^{2}}{2(3q^{2}\beta-2)},
\end{align}
and $Q^{2}+P^{2}$ is the strength of the EM field. In order for the BH event horizon to exist, one has to have $|\Gamma|<M^2$.
The asymptotic expansion coefficients of this metric are 
\begin{subequations}
    \begin{align}
        a_{1}&=-2M,\quad a_{2}=\Gamma,\quad a_n=0,~(n\geq 3)\\
        b_{1}&=2M,\quad b_{2}=4M^{2}-2\Gamma,\quad b_{3}=8M^{3}-6M\Gamma,\nonumber\\
        b_{4}&=16M^4-16M^2\Gamma+3\Gamma^2,\quad \dots
    \end{align}
\end{subequations}

Substituting these coefficients into Eqs. \eqref{eq:dphif1} and \eqref{eq:dphif2}, we obtain the deflection angle in the charged Galileon spacetime
\begin{align}
\label{eq:dphicg1}
&\Delta\phi_{\text{CG}}=\pi+\frac{2M}{b}\left( 1+\frac{1}{n_{\infty}^{2}} \right)+\frac{3\pi M^{2}}{4b ^{2}}\left[1+\frac{4}{n_{\infty}^{2}}\right.\nonumber\\
    &\left.-\frac{2\Gamma}{3M^{2}}\left( 1+\frac{1}{n_{\infty}^{2}} \right) \right]+\frac{10M^{3}}{3b^{3}}\left[ 1+\frac{9}{n_{\infty}^{2}}+\frac{3}{n_{\infty}^{4}}-\frac{1}{5n_{\infty}^{6}}\right.\nonumber\\
    &\left.-\frac{4\Gamma}{5M^{2}}\left( 1+\frac{21}{4n_{\infty}^{2}}+\frac{3}{4n_{\infty}^{4}} \right) \right]+\frac{105\pi M^{4}}{64b^{4}}\nonumber\\
    &\times\left[ 1+\frac{16}{n_{\infty}^{2}}+\frac{16}{n_{\infty}^{4}}-\frac{36\Gamma}{35M^{2}}\left( 1+\frac{34}{3n_{\infty}^{2}}+\frac{64}{9n_{\infty}^{4}} \right)\right.\nonumber\\
    &\left.+\frac{8\Gamma^{2}}{35M^{4}}\left( 1+\frac{4}{n_{\infty}^{2}}+\frac{1}{n_{\infty}^{4}} \right) \right]+\mathcal{O}\left(\frac{M^5}{b^5}\right),
\end{align}
whose small $\varepsilon$ expansion is
\begin{align}
&\Delta\phi_{\text{CG}}=\pi+\frac{4M}{b}+\frac{\pi 15(M^{2}-4\Gamma)}{4b ^{2}}+\frac{8M(16M^{2}-7\Gamma)}{3b^{3}}\nonumber\\
&+\frac{3\pi (1155M^{4}-700M^{2}\Gamma+48\Gamma^{2})}{64b^{4}}\nonumber\\
&+\varepsilon^{2}\left[ \frac{2M}{b}+\frac{\pi (6M^{2}-\Gamma)}{2b ^{2}}+\frac{3M(16M^{2}-6\Gamma)}{b^{3}}\right.\nonumber\\
&\left.+\frac{3\pi (210M^{4}-115M^{2}\Gamma+6\Gamma^{2})}{8b^{4}} \right]+\mathcal{O}\left( \varepsilon^{4}, \frac{M^{5}}{b^{5}} \right).
\end{align}

\begin{figure}[htp!]
\includegraphics[width=0.48\textwidth]{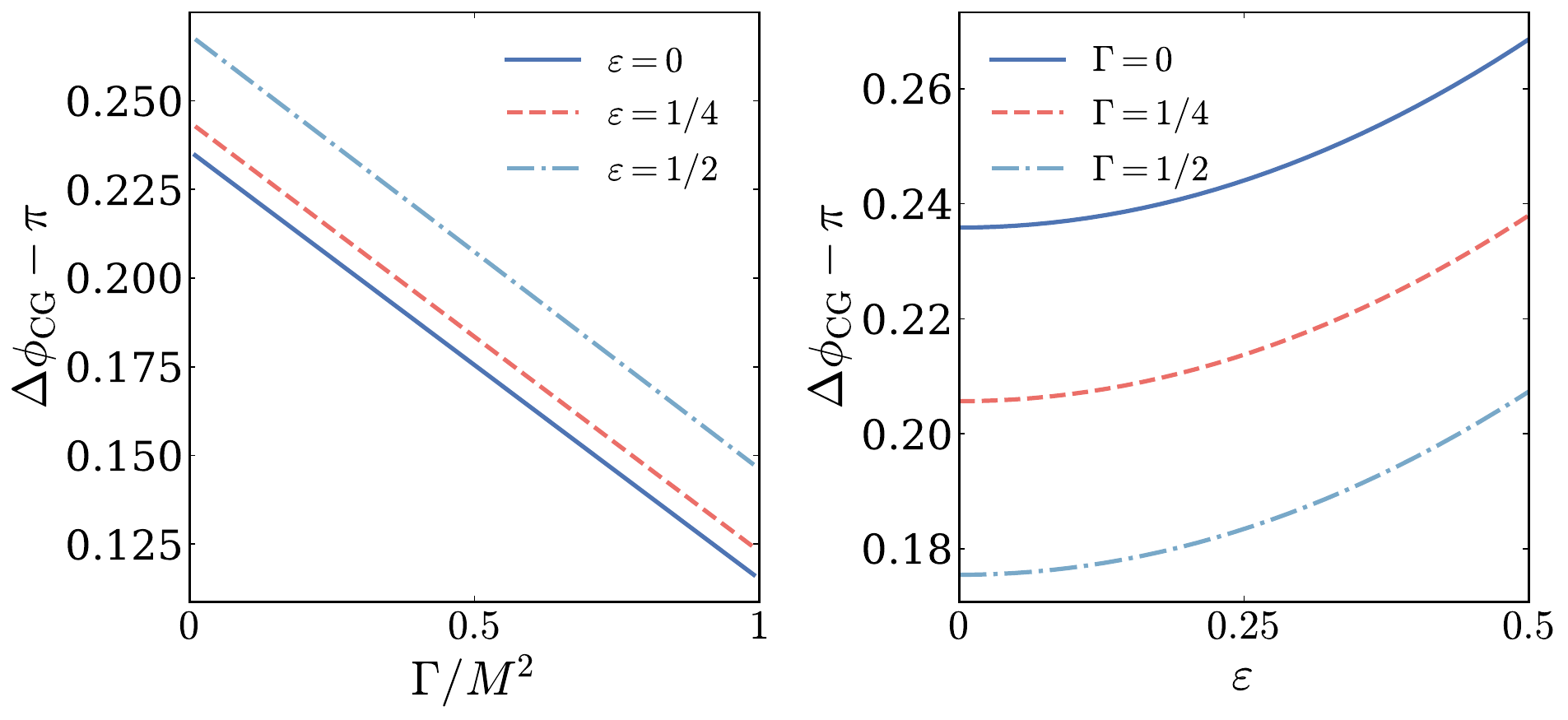}
\caption{\label{fig:cg1} Deflection angle \eqref{eq:dphicg1} as a function of $\Gamma$ (left) and $\varepsilon$ (right). 
$b=20M$ is used in plotting the figure.
}
\end{figure}

In Fig. \ref{fig:cg1} we plot the dependence of the deflection angle on the parameter $\Gamma$ and frequency ratio $\varepsilon$. It is seen that as the electromagnetic field strength $\Gamma$ increases, the deflection decreases linearly, which is also qualitatively similar to the effect of $Q^2$ in the RN spacetime. The effect of the plasma is still to increase the deflection angle, which is the same as in the RN and charged Horndeski spacetime, as stated after Eq. \eqref{eq:dphif2} for the general spacetime case.

\section{Conclusions and Discussion}\label{sec:conclusion}

In this work, we developed a perturbative technique to compute the deflection angle of light rays in the SSS spacetime within a homogeneous plasma. The result as shown in Eq. \eqref{eq:dphif1} takes the form of a series of $M/b$ with coefficients being polynomials of the asymptotic expansion coefficients of the metric functions, as well as the reciprocal of the asymptotic refractive index. We also worked out the dual series of both the small quantities $M/b$ and the frequency ratio $\varepsilon$ as shown in Eq. \eqref{eq:dphif2}. The deflection angles in Eq. \eqref{eq:dphif1} and \eqref{eq:dphif2} also reveal that the spacetime charge and the plasma have competing effects on the deflection angle. The former, to the leading order, always decreases the deflection angle, while the effect of the plasma, to the leading order of the plasma density, equals $2\varepsilon^2 M/b$, which is always positive for arbitrary SSS spacetime and therefore increases the deflection angle and the apparent angle of gravitational lensing images. 

We showed the exponential accuracy of our series results as the truncation order of the series increases, using the deflection in RN spacetime as an example. The results are applied to find the deflection angle in two other spacetimes, namely the charged Horndeski and charged Galileon spacetimes. In Appendix \ref{sec:app1}, deflection angles in previously studied spacetimes are computed using our method, and the results are compared with the corresponding earlier results. 

One great advantage of a perturbative result of the deflection angle is the additivity of the effect of various factors in their leading orders. Previously, it was known that for a counterclockwise spinning spacetime with spin angular momentum per unit mass square $a$, the deflection angle of a light ray in the equatorial plane receives a correction from spacetime spin as $\Delta\phi_a=4aM^2/b^2$ \cite{Huang:2020trl,Jia:2020xbc}. Therefore, for general stationary and axisymmetric spacetime, combining the contributions from the spacetime mass, electric charge, the spacetime spin, and the plasma, the deflection angle of light ray to the leading order of each kind of factor becomes 
\begin{align}
    \Delta\phi=\frac{4M}{b}+\varepsilon^2\frac{2M}{b}-\frac{3\pi Q^2}{b^2}+4a\frac{M^2}{b^2},\quad |a|\leq 1.
\end{align}
This formula could be used to study the effect of these factors in other phenomena based on light trajectory deflection. 

\section*{Acknowledgements}

This work is supported by the Undergraduate Training Programs for Innovation of Wuhan University (No. S202610486124).

\appendix

\section{Deflection angles in other spacetimes}
\label{sec:app1}
In this appendix, we present the deflection angles in some other spacetimes that were studied previously in literature.

\subsection{Deflection in the Einstein-Dyonic-ModMax BH spacetime}
\label{subsec:EDM_BH}

The spacetime of Einstein-Dyonic-ModMax (EDM) BH framework is described by \cite{Bokulic:2025usc}
\begin{align}
    A(r)=\frac{1}{B(r)}=1-\frac{2M}{r}+\frac{(\tilde{Q}^{2}+\tilde{P}^{2})\mathrm{e}^{-\gamma}}{r^{2}},\label{eq:EDM BH metric}
\end{align}
where $\mathrm{e}^{\gamma}$ is a constant factor of Einstein-ModMax equations in Weyl construction, $\tilde{Q}$ and $\tilde{P}$ are the electric and magnetic parts of the original charge $Q$.

Since this metric can be mapped from the RN metric \eqref{eq:rnmetric} using $Q\to (\tilde{Q}^{2}+\tilde{P}^{2})\mathrm{e}^{-\gamma}$, we can simply obtain the deflection in this spacetime by substituting the same mapping into Eq. \eqref{eq:dphirn2}. The result is 
\begin{widetext}
\begin{align}
\label{eq:EDM BH}
\Delta\phi_{\text{EDM}}=&\pi+\frac{4M}{b}+\frac{3\pi[5M^{2}-(\tilde{P}^{2}+\tilde{Q}^{2})\mathrm{e}^{-\gamma}]}{4b ^{2}}+\frac{16M[8M^{2}-3(\tilde{P}^{2}+\tilde{Q}^{2})\mathrm{e}^{-\gamma}]}{3b^{3}}\nonumber\\
&+\frac{105\pi[33M^{4}-18M^{2}(\tilde{P}^{2}+\tilde{Q}^{2})\mathrm{e}^{-\gamma}+(\tilde{P}^{2}+\tilde{Q}^{2})^{2}\mathrm{e}^{-2\gamma}]}{64b^{4}}+\varepsilon^{2}( \frac{2M}{b}+\frac{\pi [6M^{2}-(\tilde{P}^{2}+\tilde{Q}^{2})\mathrm{e}^{-\gamma}]}{2b ^{2}}\nonumber\\
&+\frac{16M[3M^{2}-(\tilde{P}^{2}+\tilde{Q}^{2})\mathrm{e}^{-\gamma}]}{b^{3}}+\frac{15\pi [42M^{4}-21M^{2}(\tilde{P}^{2}+\tilde{Q}^{2})\mathrm{e}^{-\gamma}+(\tilde{P}^{2}+\tilde{Q}^{2})\mathrm{e}^{-2\gamma}]}{8b^{4}} )+\mathcal{O}\left( \varepsilon ^{4}, \frac{M^{5}}{b^{5}} \right).
\end{align}
\end{widetext}

Eq. (45) of the article \cite{Sucu:2025dco} calculates the deflection angle in a homogeneous plasma within the EDM BH spacetime.  Its first two order terms in the Schwarzschild and no-plasma limit, i.e. terms at order $\mathcal{O}\left[\left(\frac{M}{b}\right)^1,\varepsilon^0\right]$ and $\mathcal{O}\left[\left(\frac{M}{b}\right)^2 ,\varepsilon^0\right]$ agree with our result above. However, that formula cannot reproduce the higher order Schwarzschild terms, or the leading  plasma contribution at order $\mathcal{O}\left[\left(\frac{M}{b}\right)^1,\varepsilon^2\right]$ which was known in Ref. \cite{Bisnovatyi-Kogan:2010flt,Crisnejo:2018uyn}.

\subsection{Deflection in Schwarzschild-MOG BH spacetime}\label{subsec:MOG}

The metric function of the Schwarzschild-MOG spacetime is given by \cite{Moffat:2014aja}
\begin{align}
A(r)=\frac{1}{B(r)}=1-\frac{2(1+\alpha)M}{r}-\frac{\alpha(1+\alpha)M^{2}}{r^{2}},
\end{align}\label{eq:metric:MOG}
This metric can be obtained from the RN one by the map $M\to (1+\alpha)M$, $Q\to -\alpha(1+\alpha)M^{2}$. Therefore the deflection in this spacetime with a homogeneous plasma can also be obtained by substituting the same mapping into Eq. \eqref{eq:dphirn2}. The result is
\begin{widetext}
\begin{align}
\Delta\phi_{\text{SM}}=&\pi+\frac{4(1+\alpha)M}{b}+ \frac{(15+33\alpha+18\alpha^{2})\pi M^{2}}{4b^{2}}+ \frac{(128+432\alpha+480\alpha^{2}+176\alpha^{3})M^{3}}{3b^{3}}\nonumber\\
&+\frac{(3465+15750\alpha+26565\alpha^{2}+19740\alpha^{3}+5460\alpha^{4})\pi M^{4}}{64b^{4}}+\varepsilon^{2}\left[ \frac{2(1+\alpha)M}{b}+\frac{(6+13\alpha+7\alpha^{2})\pi M^{2}}{2b ^{2}}\right.\nonumber\\
&\left.+\frac{(48+160\alpha+176\alpha^{2}+64\alpha^{3})M^{3}}{b^{3}}+\frac{(630+2835\alpha+4740\alpha^{2}+3495\alpha^{3}+960\alpha^{4})\pi M^{4}}{8b^{4}} \right]+\mathcal{O}\left( \varepsilon ^{4}, \frac{M^{5}}{b^{5}} \right). \label{eq:smres}
\end{align}
\end{widetext}

Ref. \cite{atamurotovWeakGravitationalLensing2021} also computed in its Eq. (25) the deflection angle in this spacetime with a homogeneous plasma. Its leading-order terms (proportional to $M/b$) of both the vacuum and plasma agree with Eq. \eqref{eq:smres}, but the higher-order terms proportional to $(M/b)^2$ or higher differ from our result. 

\subsection{Deflection in Kazakov–Solodukhin BH spacetime}
\label{subsec:KS_BH}

The spacetime of Kazakov–Solodukhin BH is described by \cite{Kazakov:1993ha}
\begin{align}
A(r)=\frac{1}{B(r)}=\frac{\sqrt{ r^{2}-a^{2} }}{r} -\frac{2M}{r},
\end{align}
where $r\geq a=4\sqrt{ \kappa }$ with $\kappa$ being the dimensional gravitational constant.
Expanding this asymptotically, the asymptotic coefficients are found to be
\begin{subequations}
    \begin{align}
        a_{1}&=-2M,\quad a_{2}=-\frac{a^{2}}{2} ,\nonumber\\
        a_{3}&=0 ,\quad a_{4}=- \frac{a^{4}}{8},\quad \dots\\
        b_{1}&=2M,\quad b_{2}=4M^{2}+\frac{a^{2}}{2},\quad b_{3}=8M^{3}+2a^{2}M,\nonumber\\
        b_{4}&=16M^{4}+6a^{2}M^{2}+\frac{3a^{4}}{8},\quad \dots
    \end{align}
\end{subequations}
Substituting these coefficients into Eqs. \eqref{eq:dphif2}, we obtain the deflection angle in this case 
\begin{align}
\label{eq:KS BH}
\Delta\phi_{\text{KS}}=&\pi+\frac{4M}{b}+ \frac{3\pi (10M^{2}+a^{2})}{8b^{2}}+ \frac{8M(16M^{2}+3a^{2})}{3b^{3}}\nonumber\\
&+\varepsilon^{2}\left[\frac{2M}{b}+\frac{\pi (12M^{2}+a^{2})}{4b ^{2}}+\frac{8M(6M^{2}+a^{2})}{b^{3}}\right]\nonumber\\
&+\mathcal{O}\left( \varepsilon ^{4}, \frac{M^{4}}{b^{4}} \right).
\end{align}

Ref. \cite{Javed:2021ymu} calculated the deflection angle in a homogeneous plasma in the same spacetime, with results shown in its Eq. (33). Its terms proportional to $M/b$, $\varepsilon^{2}M/b$, $a^{2}/b^{2}$ and $\varepsilon^{2} a^2/b^2$ agree with ours.

\subsection{Deflection in renormalization group improved Schwarzschild BH spacetime}
\label{subsec:RGI_BH}

The spacetime of the renormalization group improved Schwarzschild BH is described by \cite{Bonanno:2000ep}
\begin{align}
A(r)=\frac{1}{B(r)}=1-\frac{2M}{r}\left( 1+\frac{\Omega M^{2}}{r^{2}}+\frac{\Omega\gamma M^{3}}{r^{3}} \right)^{-1},
\end{align}
where the parameter $\Omega$ is a measure of the quantum gravity effects. The classical Schwarzschild limit is recovered by setting $\Omega=0$. 

The asymptotic expansion coefficients for the above metric are
\begin{subequations}
    \begin{align}
            a_{1}&=-2M,\quad a_{2}=0 ,\nonumber\\
            a_{3}&=2M^{3}\Omega,\quad a_{4}=2M^{4}\gamma\Omega,\quad \dots\\
            b_{1}&=2M,\quad b_{2}=4M^{2},\quad b_{3}=8M^{3}-2M^{3}\Omega,\nonumber\\
            b_{4}&=16M^{4}-8M^{4}\Omega-2M^{4}\gamma\Omega,\quad \dots
    \end{align}
\end{subequations}

Substituting these coefficients into Eqs. \eqref{eq:dphif2}, we obtain the deflection angle in this case as
\begin{align}\label{eq:rgi angle}
&\Delta \phi_{\text{RGI}}=\pi+\frac{4M}{b}+ \frac{15\pi M^{2}}{4b^{2}}+ \frac{128M^{3}}{3b^{3}}+\frac{3465\pi M^{4}}{64b^{4}}\nonumber\\
&+\varepsilon^{2}\left[ \frac{2M}{b}+\frac{3\pi M^{2}}{b ^{2}}+\frac{4(12-\Omega)M^{3}}{b^{3}}\right.\nonumber\\
&\left.+\frac{3(105-60\Omega-2\gamma\Omega)\pi M^{4}}{4b^{4}} \right]+\mathcal{O}\left( \varepsilon ^{4}, \frac{M^{5}}{b^{5}} \right).
\end{align}

Ref. \cite{Atamurotov:2022iwj} in its Eq. (28) gave the deflection angle in the same spacetime in a homogeneous plasma. The leading-order contribution (proportional to $M/b$) from both the vacuum and plasma agrees with Eq. \eqref{eq:rgi angle}, but the higher-order terms proportional to $(M/b)^2$ or higher differ from our result.

\subsection{Deflection in BH spacetime in Rastall theory}
\label{subsec:Rastall theory}

The metric functions of a known BH solution in Rastall theory take the following form \cite{Heydarzade:2017wxu}
\begin{align}
A(r)=\frac{1}{B(r)}=1-\frac{2M}{r}+\frac{Q^{2}}{r^{2}}- \frac{N_{\text{d}}}{r^{(1-6k\lambda)/(1-3k\lambda)}},
\end{align}
where $k$ and $\lambda$ are geometric parameters of the Rastall theory, while the integration constant $N_{\text{d}}$ is characteristic of the surrounding field. In the $N_{\text{d}}\to 0$ limit, this reduces to the metric of the RN BH spacetime. 

In order to satisfy the form of the expansion in Eq. \eqref{eq:metric expansion}, we can only deal with the case that the exponent of $r$ in the last term is an integer
\begin{align}
\frac{1-6k\lambda}{1-3k\lambda}=n,\quad (n= 3,\,4,\,\cdots)
\end{align}
from which we solve that
\begin{align}
    k\lambda=\frac{n-1}{3(n-2)}.
\end{align}
As an example, we next solve the deflection angle for $k\lambda=2/3$, i.e. $n=3$. The asymptotic expansion coefficients of the metric functions in this case are
\begin{subequations}
    \begin{align}
        a_{1}&=-2M,\quad a_{2}=Q^{2},\nonumber\\
        a_{3}&=N_{\text{d}} ,\quad a_{n}=0,\quad(n\geq 4)\\
        b_{1}&=2M,\quad b_{2}=4M^{2}-Q^{2},\quad b_{3}=8M^{3}-4M Q^{2}+N_{\text{d}},\nonumber\\
        b_{4}&=16M^{4}-12M^{2}Q^{2}+4M N_{\text{d}}+Q^{4},\quad \dots 
    \end{align}
\end{subequations}
Substituting these coefficients into Eqs. \eqref{eq:dphif2}, we obtain the deflection angle in this case 
\begin{align}
&\Delta\phi_{\text{RT}}=\pi+\frac{4M}{b}+\frac{3\pi(5M^{2}-Q^{2})}{4b ^{2}}\nonumber\\
&+\frac{16M(8M^{2}-3Q^{2})+4N_{\text{d}}}{3b^{3}}\nonumber\\
&+\frac{105\pi(33M^{4}-18M^{2}Q^{2}+Q^{4})+168\pi MN_{\text{d}}}{64b^{4}}\nonumber\\
&+\varepsilon^{2}\left[ \frac{2M}{b}+\frac{\pi (6M^{2}-Q^{2})}{2b ^{2}}+\frac{16M(3M^{2}-Q^{2})-2N_{\text{d}}}{b^{3}}\right.\nonumber\\
&\left.+\frac{15\pi (42M^{4}-21M^{2}Q^{2}+Q^{4})-48\pi MN_{\text{d}}}{8b^{4}} \right]\nonumber\\
&+\mathcal{O}\left( \varepsilon ^{4}, \frac{M^{5}}{b^{5}} \right).
\end{align}

Eq. (49) in Ref. \cite{Ali:2024cti} calculated the deflection angle in this spacetime with a plasma. However, that equation differs significantly from ours here and does not reduce to the RN spacetime result when setting $N_{\text{d}}=0$.

\bibliographystyle{unsrt}  
\bibliography{reference}  

\end{document}